\documentclass[3p,times,procedia]{elsarticle}
\usepackage{nupha_ecrc}

\volume{00}

\firstpage{1}

\journalname{Nuclear Physics A}

\runauth{A.~Kumar for the JETSCAPE Collaboration}

\jid{nupha}

\jnltitlelogo{Nuclear Physics A}

\usepackage{amssymb}
\usepackage{subcaption}

\biboptions{comma,square, compress}

\usepackage[figuresright]{rotating}

\begin{document}

\begin{frontmatter}



\dochead{XXVIIIth International Conference on Ultrarelativistic Nucleus-Nucleus Collisions\\ (Quark Matter 2019)}

\title{Jet quenching in a multi-stage Monte Carlo approach}


\author{Amit Kumar for the JETSCAPE collaboration}

\address{Department of Physics and Astronomy, Wayne State University, Detroit, MI 48201, USA}

\begin{abstract}
We present a jet quenching model within a unified multi-stage framework and demonstrate for the first time a simultaneous description of leading hadrons, inclusive jets, and elliptic flow observables which spans  multiple centralities and collision energies. This highlights one of the major successes of the JETSCAPE framework in providing a tool for setting up an effective parton evolution that includes a high-virtuality radiation dominated energy loss phase (MATTER), followed by a low-virtuality scattering dominated (LBT) energy loss phase. Measurements of jet and charged-hadron $R_{AA}$ set strong constraints on the jet quenching model. Jet-medium response is also included through a weakly-coupled transport description.
\end{abstract}

\end{frontmatter}


\section{Introduction}
\label{}
Ultra-relativistic nucleus-nucleus collisions performed at the Relativistic Heavy-Ion Collider (RHIC) and Large Hadron Collider (LHC)  produce an  exotic state of deconfined matter that  undergoes different stages,  namely: initial state hard scattering, expansion of deconfined quark-gluonic matter, jet energy loss in the medium, and hadronization. A unified framework that implements all stages of a heavy-ion collision is required to gain a comprehensive understanding of the QGP and explore new physics.  Jet Energy-loss Tomography with a Statistically and Computationally Advanced Program Envelope (JETSCAPE) is a state-of-the-art simulation framework with capabilities to accommodate physics of each stage of the heavy-ion collision in a modular form \cite{Kumar:2019bvr,Putschke:2019yrg, PhysRevC.96.024909}. The goal of these proceedings is to  demonstrate that a multi-stage jet quenching model gives a simultaneous description of the jet and hadron observables at multiple centralities and collision energies. To see JETSCAPE capabilities in studying intra-jet observables, heavy-quark observables and  Bayesian calibration of the soft sector visit Refs. \cite{YasukiJSJetMedium, GojkoHQ,JFPaquetSims}.


 \section{Multi-stage jet energy loss with JETSCAPE framework}
It has been widely accepted that
 the evolution of jets through deconfined QCD matter is a multi-scale problem. To incorporate this property within the JETSCAPE framework, we set up an effective parton evolution in which we incorporated information of the space-time evolution of the medium on parton energy loss during the high-virtuality, radiation dominated portion of the shower using MATTER, followed by a simulation of the low-virtuality, scattering dominated portion with LBT. 
 The switching between stages is performed at a parton-by-parton level depending on local quantities such as local energy density in the medium,  off-shellness and energy of the parton.

MATTER \cite{PhysRevC.88.014909,majumder2009inmedium} is a virtuality-ordered  Monte Carlo event generator that simulates the evolution of partons at high energy ($E$) and high-virtuality (off-shellness) $Q^2 \gg \sqrt{\hat{q}E} $, where $\hat{q}$ is the transport coefficient that controls transverse broadening, and $Q^2 \geq 1$ GeV$^2$. Based on the Higher-Twist formalism, the probability of a parton splitting is computed by sampling the medium modified Sudakov form factor which includes contributions from the vacuum and medium-induced splitting functions. We used a hard thermal loop formulation for $\hat{q}$ \cite{PhysRevC.82.064902}.
 The medium response is included through a weakly-coupled description of the medium in terms of thermal partons, where the propagation of the medium parton kicked out by the jet parton (``recoil") is described by a kinetic theory-based approach.

LBT \cite{PhysRevLett.111.062301, PhysRevC.91.054908} is also a Monte Carlo event generator based on the linear Boltzmann equation which simulates in-medium energy loss of high-energy, low-virtuality partons. The model includes both the elastic $2 \rightarrow 2$ scattering processes and inelastic $2 \rightarrow 2+n$ scattering with multiple gluon radiation. The probabilities of these scattering processes are employed to simulate the evolution of the jet shower, recoil partons and radiated gluons due to their scattering with thermal partons in the medium.

\begin{figure}[ht!]
  \centering
\begin{subfigure}{.4\textwidth}
  \includegraphics[width=.98\linewidth]{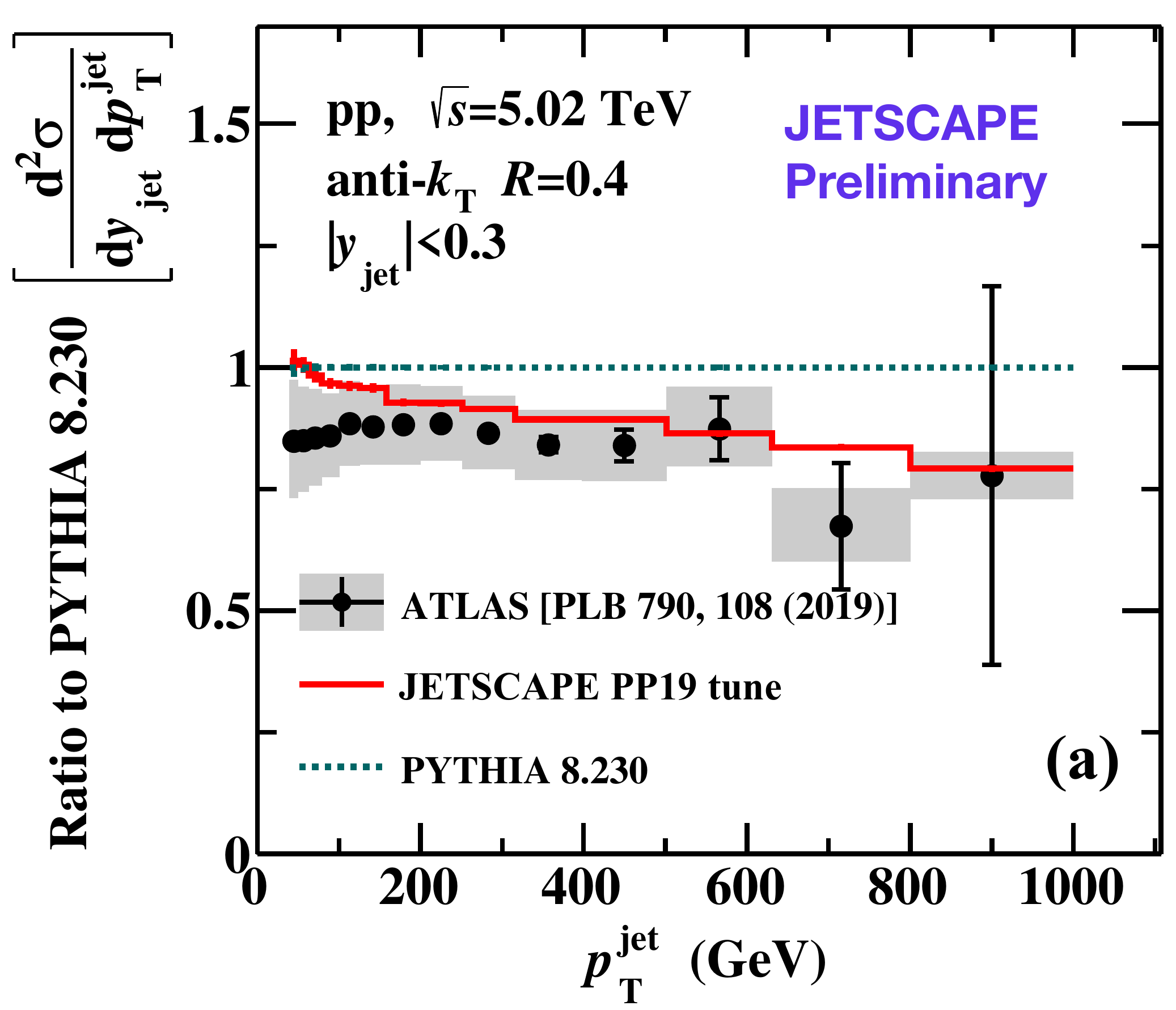}  
  
  \label{fig:sub-first}
\end{subfigure}  \hspace{0.05\textwidth}
\begin{subfigure}{.4\textwidth} 
  \includegraphics[width=.98\linewidth]{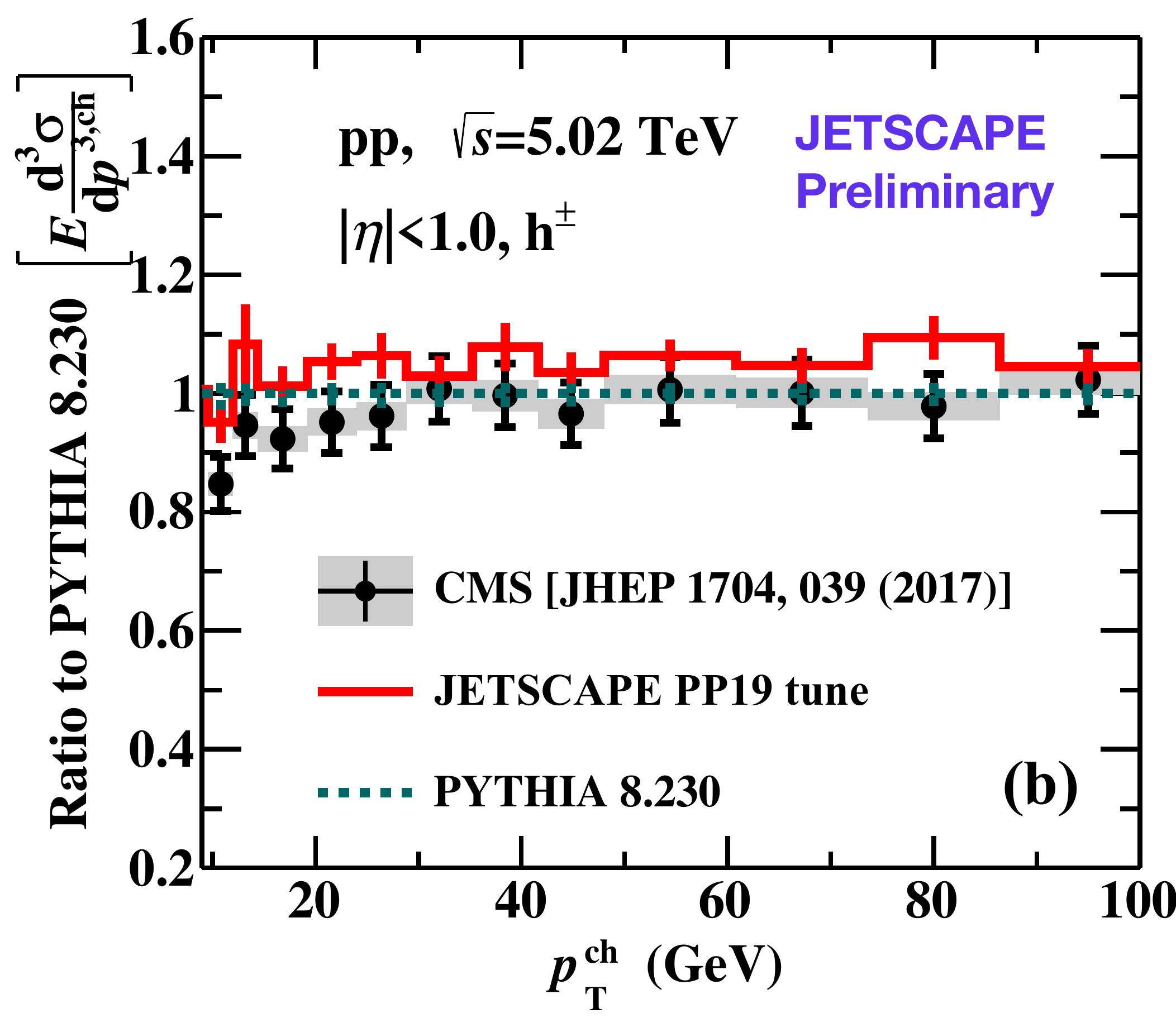}  
 
  \label{fig:sub-second}
\end{subfigure}
\caption{Comparison of JETSCAPE PP19 tune's results for inclusive jets and charged-hadrons with the experimental data and the default PYTHIA at 5.02 TeV. The ratio is taken w.r.t. to  the default PYTHIA result. (a) Ratio of inclusive jet cross-sections as a function of jet $p_{\mathrm{T}}$ for cone size $R=0.4$.  (b) Ratio of charged-hadron yield as a function of charged-hadron $p_{\mathrm{T}}$.}
\label{fig:pp_5TeV}
\end{figure}

The simulation of p+p collisions  is performed using the JETSCAPE PP19 tune \cite{Kumar:2019bvr}.  This tune employs  PYTHIA as a hard scattering module with initial state radiation (ISR), multiparton interaction (MPI) flags ON and final state radiation (FSR) flag OFF. MATTER is employed for final state radiation. Simulation of PbPb collision events at $5.02A$ TeV is done by using fluctuating initial state conditions evolved hydrodynamically within the JETSCAPE framework. TRENTO \cite{Moreland:2014oya}+PYTHIA is used to  simulate the initial state hard scattering. The medium profiles are generated using (2+1)-D VISHNU \cite{Shen:2014vra} with fluctuating TRENTO \cite{Moreland:2014oya} initial conditions. The virtuality-ordered shower is generated using an in-medium MATTER generator which evolves partons to a lower virtuality $Q_0$ (switching virtuality). Then, these partons are passed to a small-virtuality energy-loss stage, LBT.  Finally, the jet partons are hadronized using PYTHIA based string fragmentation.
  
 \section{Results}
Figure \ref{fig:pp_5TeV} displays a ratio of inclusive jet cross-sections (left) and charged-hadron yields (right). 
Overall, the JETSCAPE  results are compatible with the experimental data (within $ 10 \%$). Results from PYTHIA tend to be similar to JETSCAPE  for the case of the charged-hadron yield but for jets show discrepancies  of order  $\lesssim 25 \%$. Both observables provide a good baseline for their counterparts in nuclear collisions. 
  
 \begin{figure}[ht!]
\begin{subfigure}{.47\textwidth}
  \centering
  \includegraphics[width=.998\linewidth]{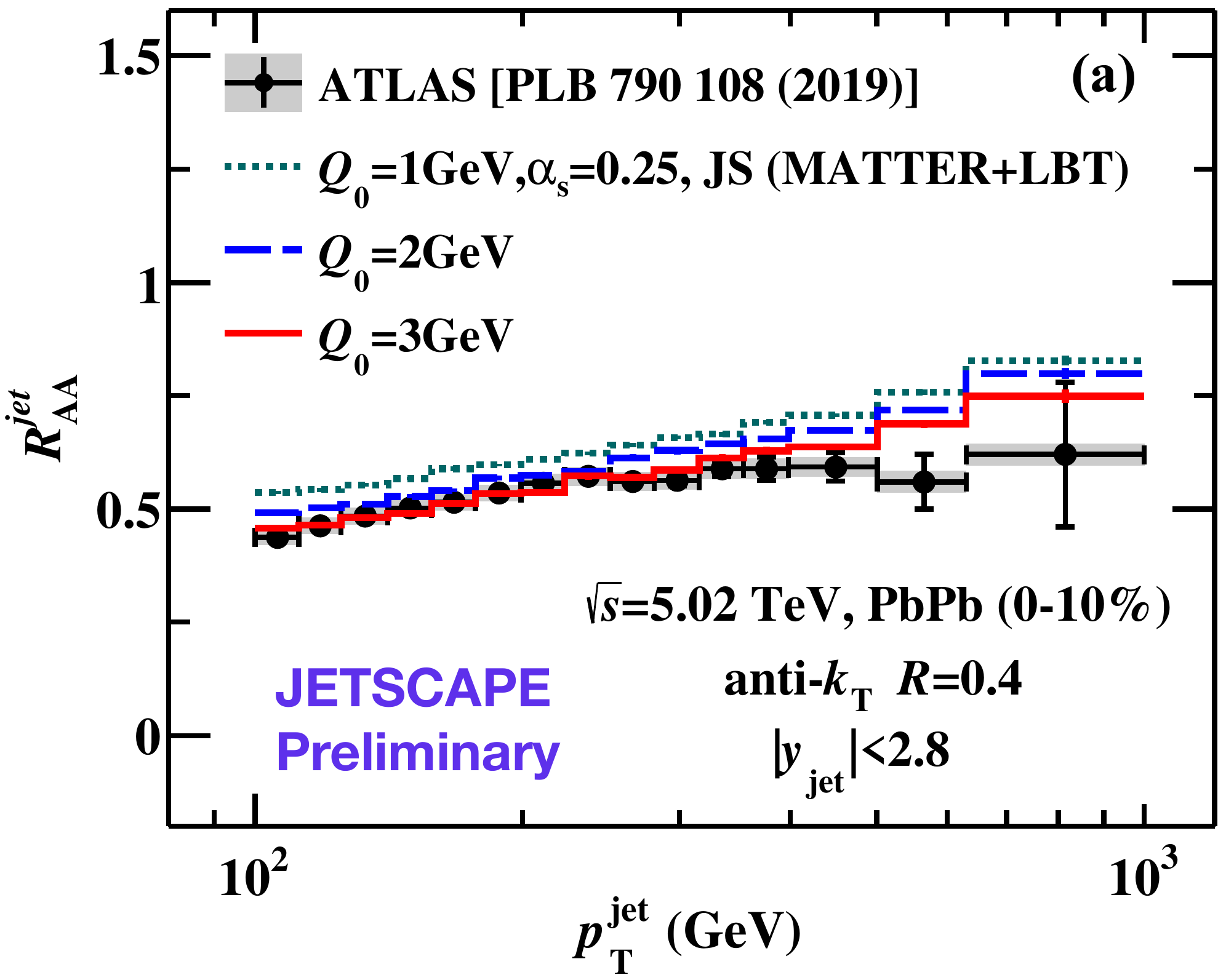}  
  
  \label{fig:sub-first}
\end{subfigure}
\begin{subfigure}{.47\textwidth}
  \centering
  \includegraphics[width=.94\linewidth]{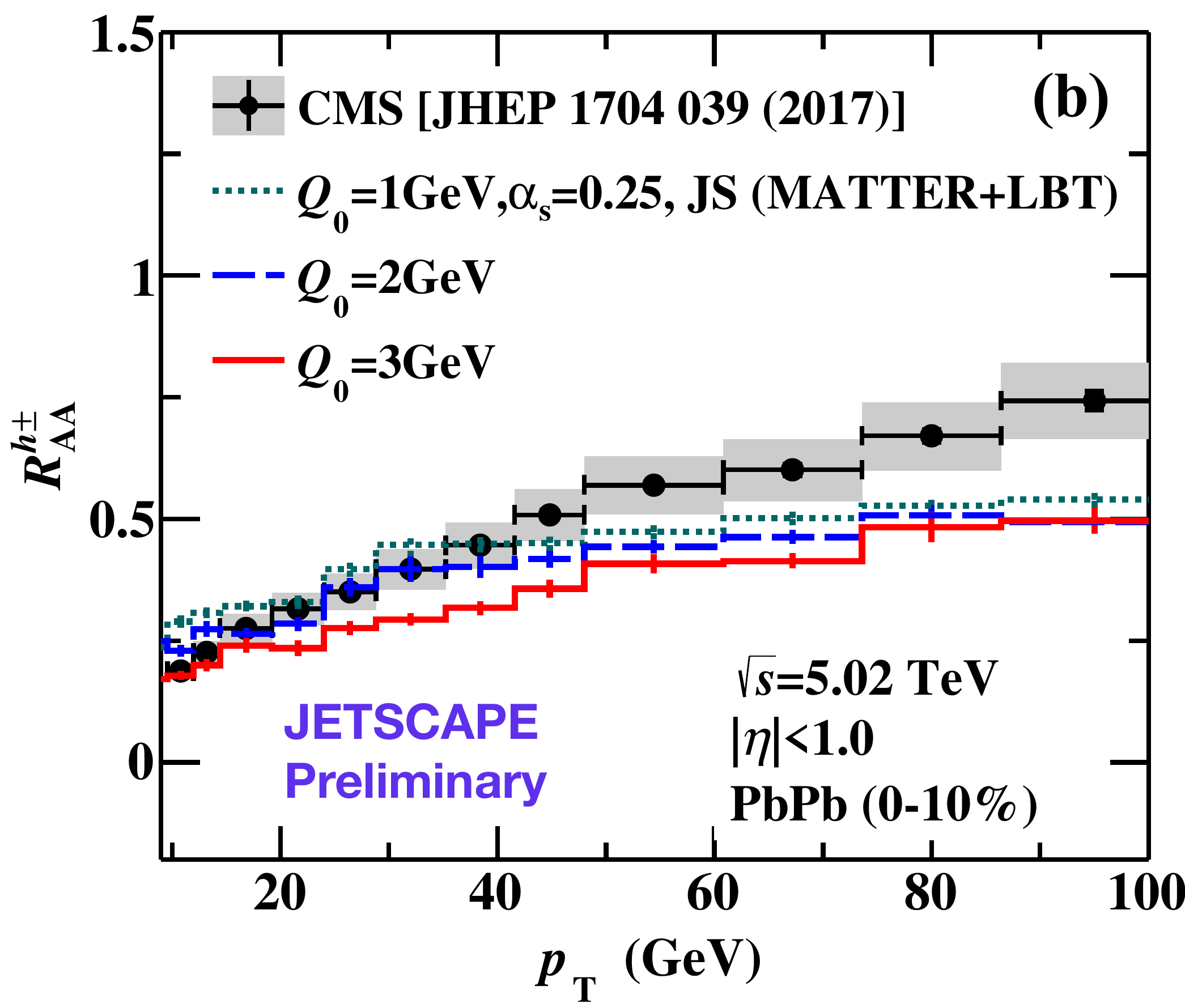}  
 
  \label{fig:sub-second}
\end{subfigure}
\caption{Comparison of the nuclear modification factor for inclusive jets and charged-hadrons obtained using the multi-stage energy loss approach (MATTER+LBT) within the JETSCAPE framework, with the experimental data for most central $5.02A$ TeV  collisions (0-10$\%$). Results are shown for   a switching virtuality parameter $Q_0$ = 1, 2, and 3 GeV. (a) Jet  $R_{AA}$ as a function of jet $p_{\mathrm{T}}$ for cone size $R=0.4$.  (b) Charged-hadron $R_{AA}$ as a function of charged-hadron $p_{\mathrm{T}}$.}
\label{fig:RAA_central_5TeV}
\end{figure}

 \begin{figure}[ht!]
\begin{subfigure}{.33\textwidth}
  \centering
  \includegraphics[width=.99\linewidth]{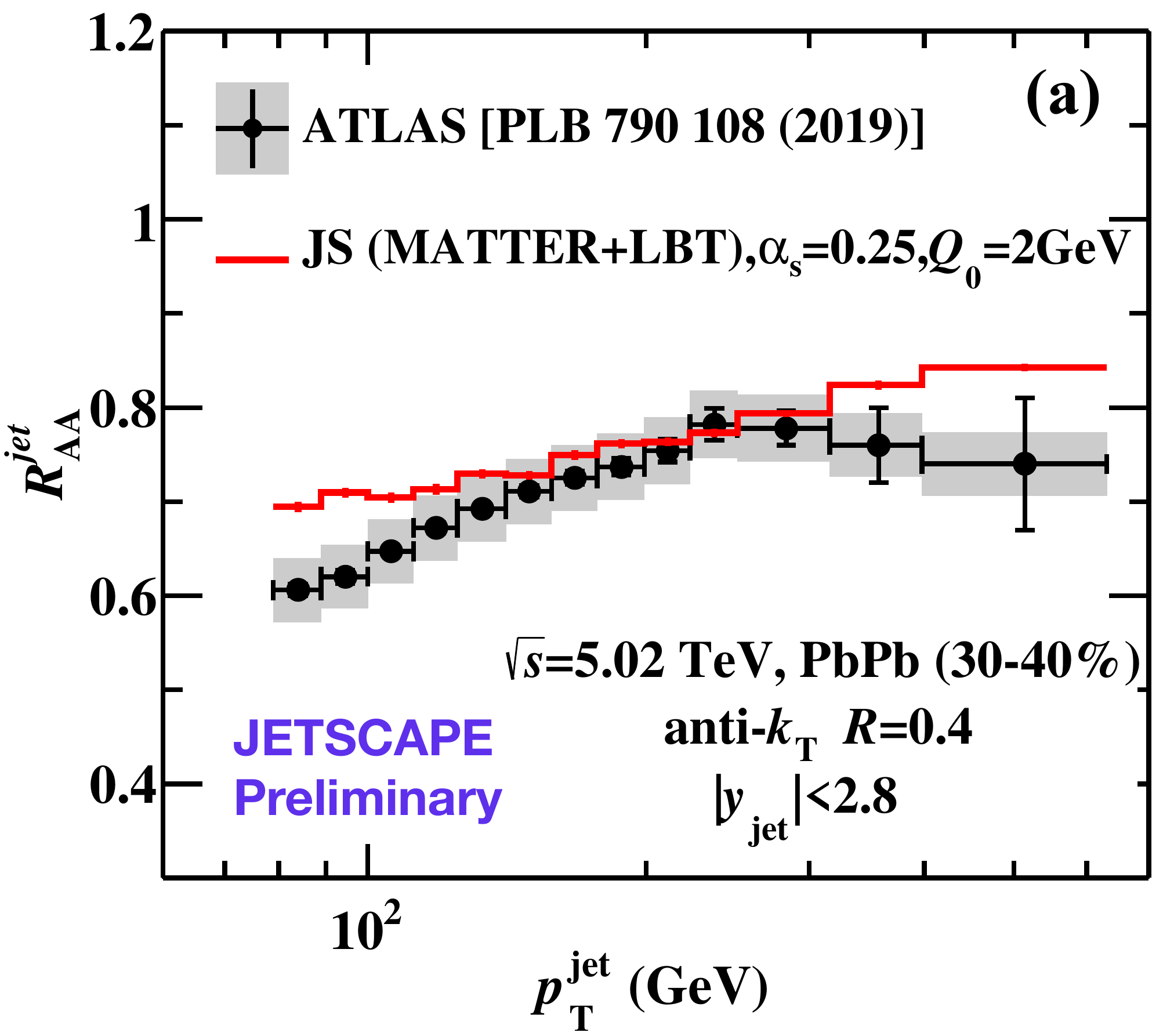}  
  
  \label{fig:sub-first}
\end{subfigure}
\begin{subfigure}{.33\textwidth}
  \centering
  \includegraphics[width=.99\linewidth]{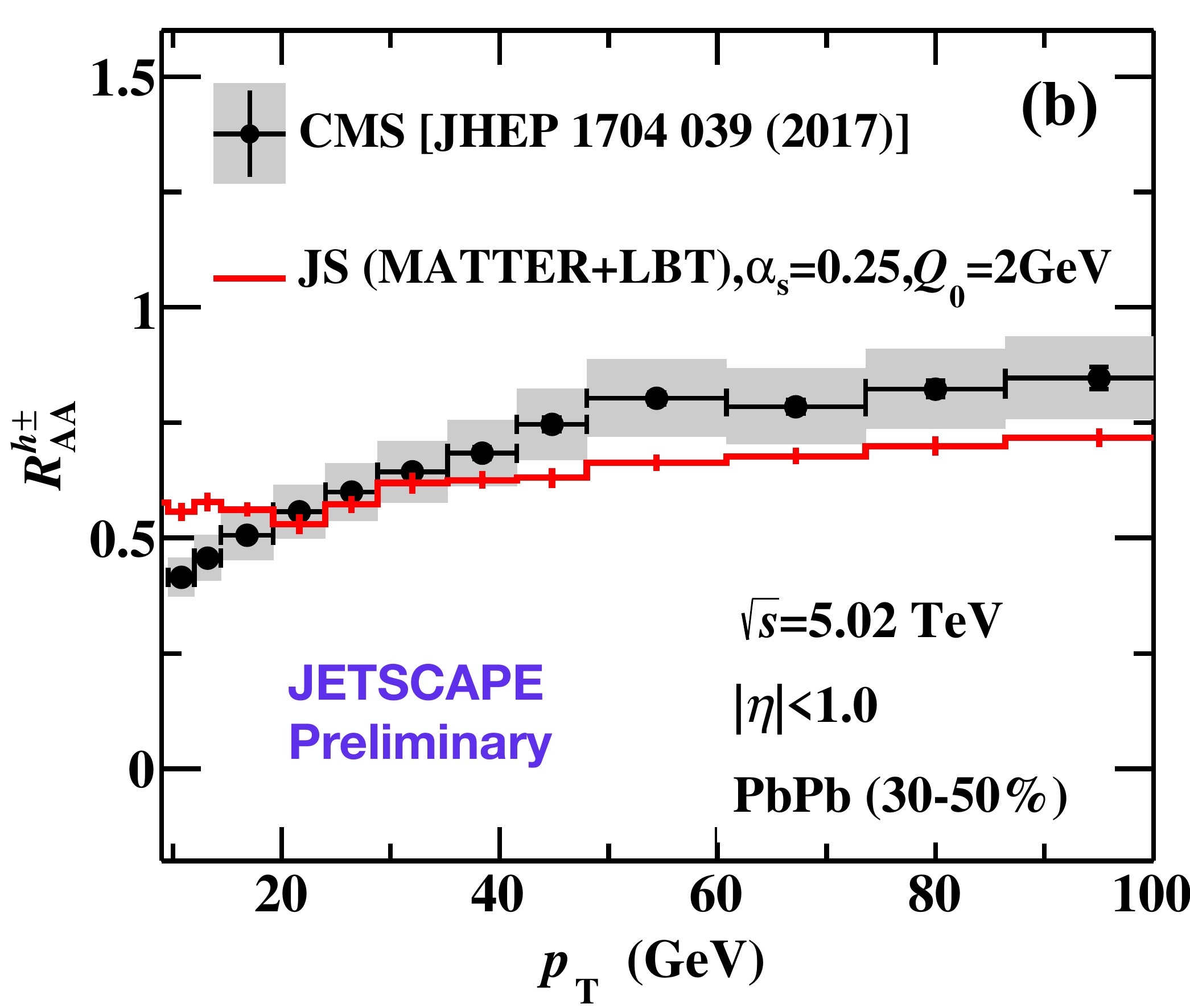}  
  
  \label{fig:sub-second}
\end{subfigure}
\begin{subfigure}{.33\textwidth}
  \centering
  \includegraphics[width=.99\linewidth]{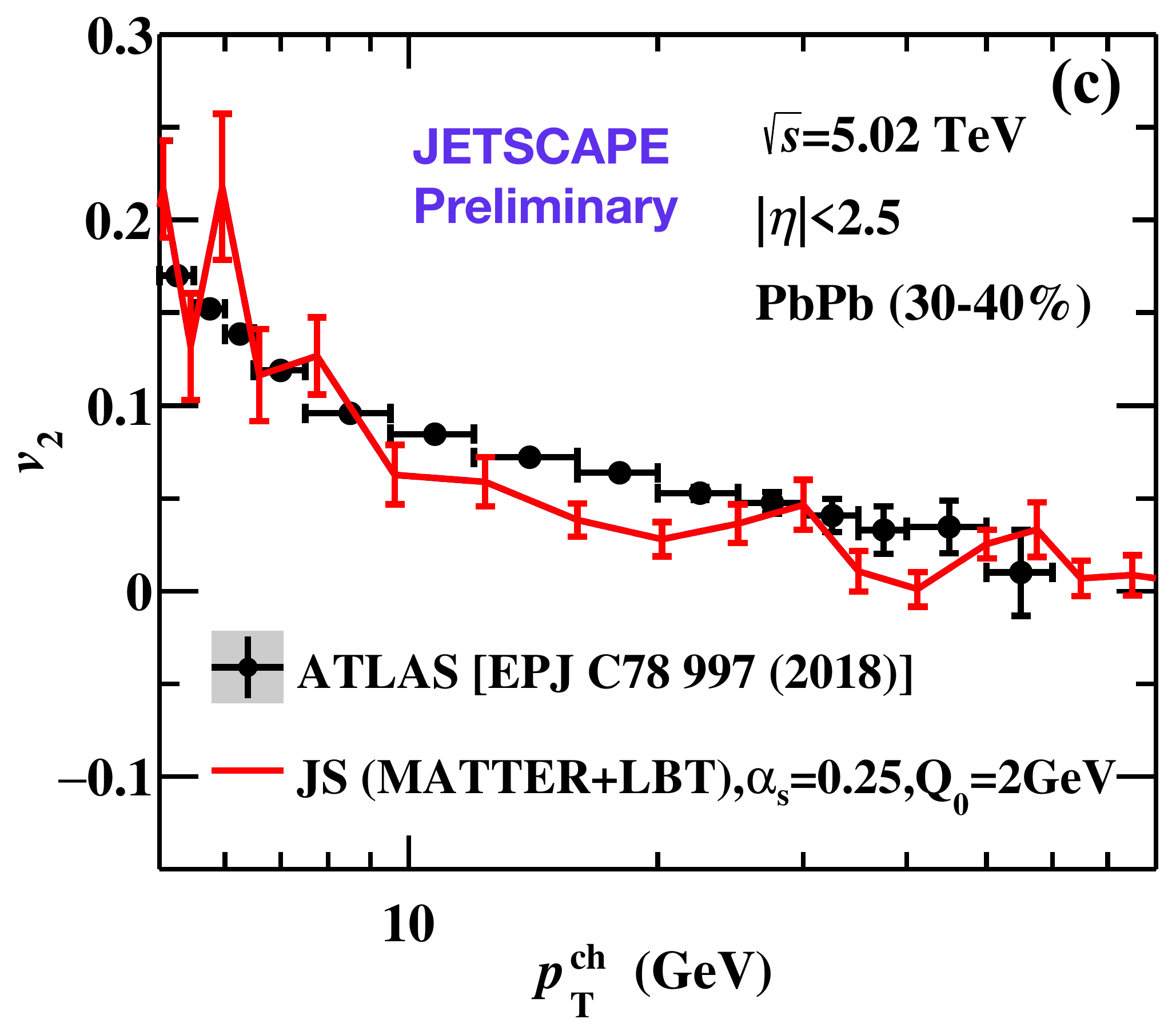}  
 
  \label{fig:sub-first}
\end{subfigure}
\caption{Comparison of inclusive jet and charged-hadron nuclear modification factor, and azimuthal anisotropy $\nu_2$ obtained using the multi-stage energy loss approach (MATTER+LBT) within the JETSCAPE framework, with the experimental data  for semi-peripheral $5.02A$ TeV collisions. The tuning parameters $Q_0$ and $\alpha_s$ are set by the simultaneous fit of inclusive jet  and charged-hadron $R_{AA}$ in the most central $5.02A$ TeV collisions. (a) Jet  $R_{AA}$ as a function of jet $p_{\mathrm{T}}$ for cone size $R=0.4$.  (b) Charged-hadron $R_{AA}$ as a function of charged-hadron $p_{\mathrm{T}}$. (c) Azimuthal anisotropy $\nu_2$ for charged-hadrons as a function of charged-hadron $p_{\mathrm{T}}$.}
\label{fig:RAAandV2_peripheral_5TeV}
\end{figure} 
 
Figure \ref{fig:RAA_central_5TeV} shows inclusive jet $R_{AA}$ (left) and charged-hadron $R_{AA}$ (right) in most central $5.02A$ TeV PbPb collisions (0-10$\%$) for switching virtualities $Q_0=1,2,3$ GeV and the jet-medium coupling parameter $\alpha_s=0.25$. Increasing $Q_0$ from 1 to 3 GeV increases the effective length of LBT based energy-loss. Since the partons in the LBT stage are close to on-shell ($Q_0 \sim [1,3]$ GeV, $E \sim p_{T}$), the partons at low-$p_{\mathrm{T}}$ see significant  energy loss effects. Our results are consistent with this concept. Increasing $Q_0$ from 1 to 3 GeV suppresses the low-$p_{\mathrm{T}}$ region of the charged-hadron $R_{AA}$ spectrum. This leads to suppression of the jet $R_{AA}$ at all jet $p_{\mathrm{T}}$'s. The parameter set $Q_0=2$ GeV and $\alpha_s=0.25$ provides the best simultaneous description of the jet and hadron $R_{AA}$ data. We observe a $\sim20\%$ deviation at high $p_{\mathrm{T}}$ in the  charged-hadron $R_{AA}$ that could be due to the absence of any scale dependence (hard parton's off-shellness) in
 $\hat{q}$ \cite{Kumar:2019uvu}. 
 
 \begin{figure}[ht!]
\begin{subfigure}{.51\textwidth}
  \centering
  \includegraphics[width=.998\linewidth]{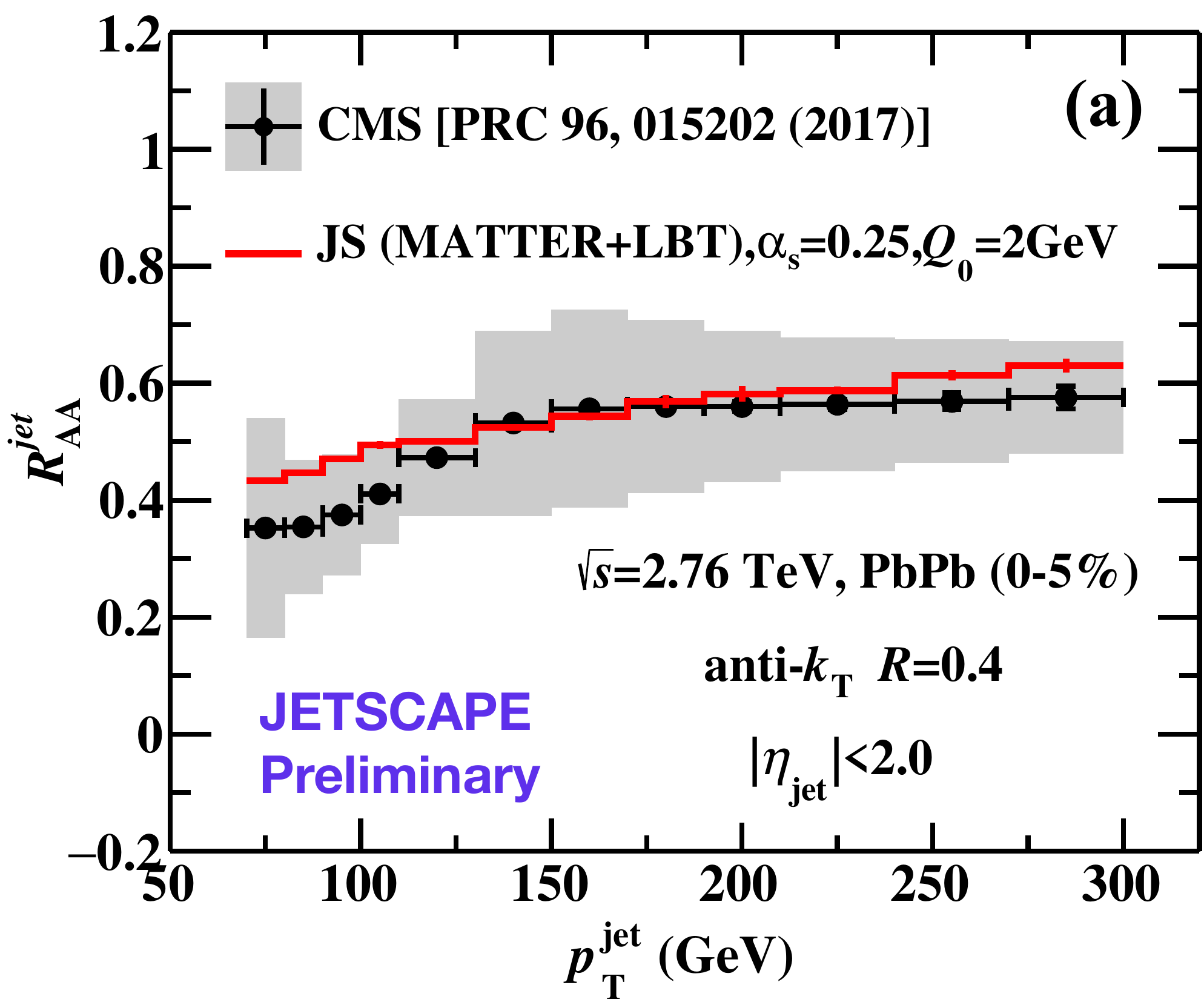}  
  
  \label{fig:sub-first}
\end{subfigure}
\begin{subfigure}{.49\textwidth}
  \centering
  \includegraphics[width=.96\linewidth]{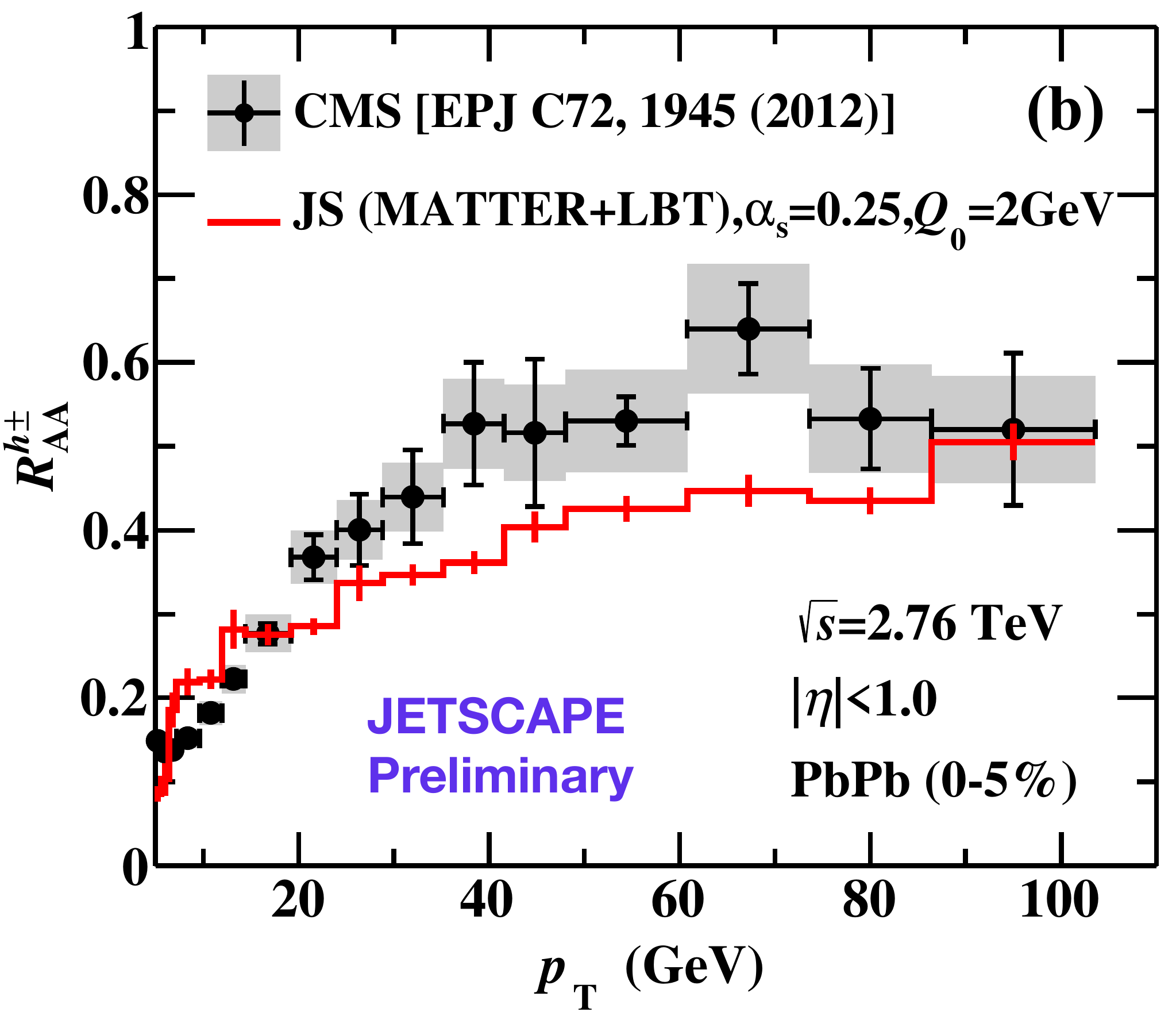}  
  
  \label{fig:sub-second}
\end{subfigure}
\caption{Comparison of the nuclear modification factor for  inclusive jets and charged-hadrons obtained using multi-stage energy loss approach (MATTER+LBT) within the JETSCAPE framework with the experimental data for most-central $2.76A$ TeV  collisions. The tuning parameters $Q_0$ and $\alpha_s$ are set by the simultaneous fit of the inclusive jet  and charged-hadron $R_{AA}$ in the most central $5.02A$ TeV collisions. (a) Jet  $R_{AA}$ as a function of jet $p_{\mathrm{T}}$ for cone size $R=0.4$.  (b) Charged-hadron $R_{AA}$ as a function of charged-hadron $p_{\mathrm{T}}$.}
\label{fig:RAA_central_2TeV}
\end{figure}

 Figure \ref{fig:RAAandV2_peripheral_5TeV} shows inclusive jet $R_{AA}$ (left), charged-hadron $R_{AA}$ (center) and azimuthal anisotropy coefficient $\nu_2$ (right) in semi-peripheral $5.02A$ TeV PbPb collisions for the parameter set ($Q_0$ and $\alpha_s$) extracted from fits to the most central  $5.02A$ collisions inclusive jet $R_{AA}$ and charged-hadron $R_{AA}$ data. The agreement with the experimental measurement is within $10\%$.

  Figure \ref{fig:RAA_central_2TeV} shows inclusive jet $R_{AA}$ (left) and charged-hadron $R_{AA}$ (right) in the most central $2.76A$ TeV PbPb collisions for the parameter set ($Q_0$ and $\alpha_s$)  extracted from fits to the most central collision inclusive jet $R_{AA}$ and charged-hadron $R_{AA}$ data at $5.02A$ TeV. Results are consistent with the experimental measurements. 

 \section{Conclusion}
 A  multi-stage energy-loss based jet quenching model  is presented. This unified approach effectively captures the physics of multi-scale jet quenching in QCD plasma. 
The free parameters in pp collisions are fixed by the JETSCAPE PP19 tune. A simultaneous fit of the jet $R_{AA}$ and charged-hadron $R_{AA}$ in the most central $5.02A$ TeV collisions has been carried out to constrain the jet quenching model. The extracted parameters  describe charged-hadron and jet $R_{AA}$, and the elliptic flow coefficient $\nu_2$ in  semi-peripheral collisions at multiple collision energies well without further re-tuning the fit parameters.

 {$ \mathbf {Acknowledgements:}$} The work is  supported in part by the US National Science Foundation (NSF), under grant numbers ACI-1550300. 



\bibliographystyle{elsarticle-num}
\bibliography{ref.bib}







\end{document}